\input harvmac
\noblackbox
\def\mnbox#1#2{\vcenter{\hrule \hbox{\vrule height#2in
                \kern#1in \vrule} \hrule}}  
\def\sq{\,\raise.5pt\hbox{$\mnbox{.09}{.09}$}\,}
\def\sqb{\,\raise.5pt\hbox{$\overline{\mnbox{.09}{.09}}$}\,}


\def\half{{1\over 2}}

\def\a{{\alpha}}

\def\d{{\delta}}

\def\CC{{\cal C}}
\def\CN{{\cal N}}

\def\cabc{\langle \CC^{I_1} \CC^{I_2} \CC^{I_3} \rangle}

%

%
%
\def\eqnn#1{\xdef #1{(\secsym\the\meqno)}\writedef{#1\leftbracket#1}%
\global\advance\meqno by1\wrlabeL#1}
\def\eqna#1{\xdef #1##1{\hbox{$(\secsym\the\meqno##1)$}}
\writedef{#1\numbersign1\leftbracket#1{\numbersign1}}%
\global\advance\meqno by1\wrlabeL{#1$\{\}$}}
\def\eqn#1#2{\xdef #1{(\secsym\the\meqno)}\writedef{#1\leftbracket#1}%
\global\advance\meqno by1$$#2\eqno#1\eqlabeL#1$$}
\Title
{\vbox{
\baselineskip12pt
\hbox{DFUB 99-12}\hbox{hep-th/9907047}}}
{\vbox{
\centerline{Three Point Functions of Chiral Primary Operators}
\centerline{in $d=3$, $\CN= 8$ and $d=6$, $\CN=(2,0)$ SCFT} 
\centerline{at Large $N$} }} 
\centerline{Fiorenzo Bastianelli and  Roberto Zucchini}
\centerline{\it Dipartimento di Fisica, Universit\`a degli Studi di Bologna}
\centerline{\it via Irnerio 46, I-40126 Bologna, Italy}
\centerline{and}
\centerline{\it I. N. F. N., Sezione di Bologna}
\vskip1cm          
\centerline{\bf Abstract}
\medskip
\noindent 

We use the $AdS$/CFT correspondence to calculate 
three point functions of chiral primary operators at large $N$
in $d=3$, $\CN= 8$ and  $d=6$, $\CN=(2,0)$ superconformal 
field theories. These theories are related to the infrared 
fixed points of world-volume descriptions of $N$ coincident 
M2 and M5 branes, respectively.

The computation can be generalized by employing a gravitational action 
in arbitrary dimensions $D$, coupled to a $(p+1)$-form and appropriately 
compactified on $AdS_{(D-p-2)} \times S^{(p+2)}$.
We note a surprising coincidence: this generalized model reproduces for
$D=10$, $p=3$ the three point functions of $d=4$, 
$\CN=4$ SYM chiral primary operators at large $N$.

\vskip 0.5cm
\Date{}

The $AdS$/CFT correspondence 
\ref\M{
J. Maldacena, ``The Large $N$ limit of Superconformal Field Theories 
and Supergravity'', Adv. Theor. Math. Phys. {\bf 2} (1998) 231, 
{\tt hep-th/9711200.}}\ref\GKP{S. Gubser, 
I. Klebanov and A. Polyakov, ``Gauge Theory
Correlators from Non-critical String Theory,'' 
Phys. Lett. {\bf B428} (1998) 105, 
{\tt hep-th/9802109}.}\ref\W{E. Witten, 
``Anti-de Sitter Space and Holography'',
Adv. Theor. Math. Phys. {\bf 2} (1998) 253, {\tt hep-th/9802150}.}
has been quite useful in learning 
various properties of strongly coupled quantum fields theories 
at large $N$ (for reviews see 
\ref\JLP{J.L. Petersen,
``Introduction to the Maldacena Conjecture on AdS/CFT'',
{\tt hep-th/9902131}.}\ref\AGMOO{
O. Aharony, S. Gubser, J. Maldacena, H. Ooguri and Y. Oz,
``Large $N$ Field Theories, String Theory and Gravity'',
{\tt hep-th/9905111}.}.)
In particular, it has been employed in
\ref\LMRS{ 
S. Lee, S. Minwalla, M. Rangamani and N. Seiberg,
``Three-Point Functions of Chiral Primary Operators in $D=4$, $N=4$ SYM 
at Large $N$'', Adv. Theor. Math. Phys. {\bf 2} (1998) 697, 
{\tt hep-th/9806074}.} 
to compute two and three point functions at large $N$ for 
chiral primary operators (CPO) of $d=4$, $\CN=4$ super Yang-Mills (SYM) 
theory by making use of type IIB supergravity compactified on
$AdS_5 \times S^5$.

This interesting result has been generalized in
\ref\CFM{ 
R. Corrado, B. Florea and R. McNees, 
``Correlation Functions of Operators and Wilson Surfaces
in the $d=6$, $(0,2)$ Theory in the Large $N$ Limit'',
{\tt hep-th/9902153}.} to the case of the $d=6$, $\CN=(2,0)$ 
superconformal field theory (SCFT) related to the infrared fixed point 
of the world-volume description of $N$ coincident M5 branes.
This is a remarkable theory which seems to require a generalization
of non-abelian gauge theories, where the gauge potential should be
though of a two-form with self-dual field strength.
A lagrangian formulation of this theory is not known, yet, but there 
exists a DLCQ Matrix description as quantum mechanics 
on the moduli space of instantons 
\ref\ABKSS{
O. Aharony, M. Berkooz, S. Kachru, N. Seiberg and E. Silverstein,
``Matrix Description of Interacting Theories in Six Dimensions'',
Adv. Theor. Math. Phys. {\bf 1} (1998) 148, {\tt hep-th/9707079}\semi
O. Aharony, M. Berkooz and N. Seiberg, ``Light-Cone Description
of $(2,0)$ Superconformal Theories in Six Dimensions'',
Adv. Theor. Math. Phys. {\bf 2} (1998) 119, {\tt hep-th/9712117}.}. 
The $AdS$/CFT conjecture has provided new
clues on this somewhat unaccessible theory.

Both theories described above have maximal supersymmetry, 
corresponding to the existence of 16 real supercharges
(for a review see \ref\S{N. Seiberg,
``Notes on Theories with 16 Supercharges'',
Nucl. Phys. Proc. Suppl. {\bf 67} (1998) 158, {\tt hep-th/9705117}.}
and references therein).
In this letter we complete the program of computing 
two and three point functions  of chiral primary operators at large $N$
for the remaining known maximally supersymmetric conformal field theory:
the $d=3$, $\CN= 8$ SCFT. This theory can be realized as the 
infrared fixed point of $N$ coincident M2 branes.
It is a quite mysterious theory without an explicitly known 
lagrangian realization,
though it can be seen to describe the strongly coupled infrared 
fixed point of $d=3$, $\CN=8$ SYM.
A kind of DLCQ Matrix  description is also unknown,
and the $AdS$/CFT conjecture offers a unique chance to learn
more about its properties.
With this aim in mind, we set out to compute the three point functions
of chiral primary operators.
For this purpose the $AdS$/CFT correspondence instruct us to:  
$i)$ identify the Kaluza-Klein tower of scalar excitations with lowest mass
arising from 11$D$ supergravity on $AdS_4 \times S^7$,
$ii)$  compute the quadratic and cubic part of their action.
This will produce the correlation functions of interest
by using the $AdS$/CFT relations. The conformal dimensions of the 
chiral primary operators have already been identified by using
Maldacena's conjecture in
\ref\confdim{O. Aharony, Y. Oz and Z. Yin, 
``M Theory on ${\rm AdS}_p\times {\rm S}_{11-p}$ and Superconformal Field
Theories'', Phys. Lett. {\bf B430} (1998) 87, {\tt hep-th/9803051}\semi
S. Minwalla, ``Particles on {$AdS_{4/7}$} and Primary Operators 
on {$M_{2/5}$}-Brane World Volumes'', 
J. High Energy Phys. {\bf 10} (1998) 002, {\tt hep-th/9803053}\semi
E. Halyo, 
``Supergravity on {$AdS_{4/7}\times S^{7/4}$} and  M  Branes'', 
J. High Energy Phys. {\bf 04} (1998) 011, {\tt hep-th/9803077}.}.

We start from the bosonic part of the 11$D$ supergravity action which 
is given by
\eqn\u{
S= {1\over 2 \kappa^2} \int d^{11}x\ 
\biggl [ \sqrt{- g}
\Bigl({1\over 2}R -{1\over {48}} F^2 \Bigr) + {1\over 3 \sqrt 2} 
A \wedge F \wedge F \biggr]}
where $F=dA_3$ is the field strength of the $3$-form $A_{mnp}$.
One can check that on the $AdS_4 \times S^7$ background the 
Chern-Simon term is never excited by the quadratic and cubic fluctuations 
of the scalar field of interest, which we denote by $s$.
Thus, for our purposes, the Chern-Simon term can be dropped from the beginning.
At this stage we also notice that a more convenient dual formulation
with a $6$-form $A_6$, instead of the original $3$-form $A_3$, can be used.
As shown in \ref\BZ{F. Bastianelli and R. Zucchini,
``Bosonic Quadratic Actions for 11-$D$ Supergravity on  
$AdS_{7/4}\times S^{4/7}$'', {\tt hep-th/9903161}.},
this has the advantage that the scalar field $s$ 
is suitably described off-shell by a linear combination of two scalars 
(namely, the deformation of the trace of the metric on the sphere
and a scalar deformation of the $6$-form potential on the sphere;
in the original $3$-form formulation this second scalar excitation
would have been described off-shell by a vector field).

The proposed simplified model captures all of the physical information
we need to extract from 11$D$ supergravity. 
It is also evident that one can generalize it
by considering a gravitational action in arbitrary dimensions  $D$ 
with metric $g_{mn}$ coupled to a $(p+1)$-form $A_{m_1.. m_{p+1}}$
\eqn\d{
S= {1\over 2 \kappa^2} \int d^{D}x\ \sqrt{- g}
\biggl [{1\over 2}R -{1\over {2 (p+2)!}} F^2 \biggr],
}
where $F= d A_{p+1}$ is the $p+2$ form describing the field strength.
For $D=11$, $p=5$ this gives the dual formulation described above, 
and constitutes the main object of our analysis.
For $D=11$, $p=2$ it gives the usual description of
11$D$ supergravity with fermions and Chern-Simon term set to zero.

The equation of motion obtained from eq. \d\ are
\eqn\t{\eqalign {&R_{mn} =  {1\over {(p+1)!}}F_{m...} F_n{}^{...}
-{(p+1)\over {(p+2)! (D-2)}} g_{mn}F^2, \cr
& \nabla^n F_{n m_1 .. m_{p+1} } = 0. \cr}}
They are easily seen to admit the 
$AdS_{(D-p-2)}\times S^{(p+2)}$ solution
induced by a Freund-Rubin type of ansatz for the 
$(p+2)$-rank antisymmetric tensor field $F$ 
\eqn\q{\eqalign{ 
F \ {\rm field\ strength}: \ \ \ \ 
&F_{\alpha_1 .. \alpha_{p+2}}
= e \epsilon_{\alpha_1 .. \alpha_{p+2}},
\  \ {\rm other}\ F =0, \cr
AdS_{(D-p-2)}:\ \ \ \
& R_{\mu\nu\lambda\rho }=- a_1
(g_{\mu\lambda} g_{\nu\rho} - g_{\mu\rho}g_{\nu\lambda}), \  \ 
a_1 = {(p+1)\over{(D-2) (D-p-3)}} e^2 ,
\cr
S^{(p+2)}: \ \ \ \ 
& R_{\alpha\beta\gamma\delta}= a_2 (g_{\alpha\gamma}
g_{\beta\delta}- g_{\alpha\delta}g_{\beta\gamma}),\ \
a_2 = {(D-p-3)\over{(p+1)(D-2)}} e^2,
\cr}}
where $\epsilon_{\alpha_1 .. \alpha_{p+2}}$ denotes the standard volume form
on the sphere and $e$ is an arbitrary mass scale parameterizing 
the compactification.
Note the splitting of $D$-dimensional 
coordinates $x^m = (z^\mu, y^\alpha) \sim (AdS_{(D-p-2)}, S^{(p+2)})$.
Also, it will be  convenient to use the notation $\bar e^2 = a_2$,
since $\bar e$ gives directly the inverse radius of the sphere.
The $(p+1)$-form $A_{p+1}$ couples naturally to ``electric''
$p$-branes, though the solution just described is generated by 
``magnetic'' $(D-p-4)$-brane sources. In fact, the $F_{p+2}$
field strength has non-trivial flux on the sphere $S^{(p+2)}$
surrounding the $(D-p-4)$-dimensional source.

Bulk fields of interest are fluctuations about the 
$AdS_{(D-p-2)}\times S^{(p+2)}$ background.
For our needs, we define them as follows (now we denote 
field variables as $g_{mn}, A_{m_1 ..m_{p+1}}, F_{m_1 ..m_{p+2}}$ 
and their background values as $\bar g_{mn}, \bar A_{m_1 ..m_{p+1}},
\bar F_{m_1 ..m_{p+2}}$ ):
\eqn\sei{\eqalign{
& h_{mn} = g_{mn} - \bar g_{mn},\ \ \ \
 h_2 = \bar g^{\alpha\beta}h_{\alpha\beta}, \ \ \ \ 
H_{\mu\nu}=h_{\mu\nu}+ {1\over {(D-p-4)}} \bar g_{\mu\nu} h_2 ,\cr
&  \delta F_{m_1 .. m_{p+2}} = 
F_{m_1 .. m_{p+2}}  -\bar{F}_{m_1 .. m_{p+2}},\ \ \
\delta F_{m_1 m_2 .. m_{p+2}}= (p+2) \bar 
\nabla_{[m_1}a_{m_2 ..  m_{p+2}]},\cr
& e a_{\alpha_1 ..  \alpha_{p+1}} = -
\epsilon_{\beta \alpha_1 .. \alpha_{p+1}}
\bar \nabla^\beta b.\cr
}} 
In the following we will drop  bars on background fields, since no 
confusion can arise. Last equation defines the scalar field $b$. 
It is the unique physical 
fluctuation contained in $a_{\alpha_1 ..  \alpha_{p+1}}$,
since we have chosen to fix reparametrization and $(p+1)$-form gauge
invariance on the sphere by imposing the gauge choices
$\nabla^\alpha h_{(\alpha \beta)}
=\nabla^\alpha h_{\alpha \mu} = \nabla^\alpha a_{\alpha m_2.. m_{p+1}}=0$, 
where $h_{(\alpha\beta)}$ denotes the traceless part of $h_{\alpha\beta}$.
We use the notation $ \sq_1 \equiv \nabla_\mu \nabla^\mu$ and  
 $ \sq_2 \equiv \nabla_\alpha \nabla^\alpha $ to denote the 
d'alembertian on $AdS$ and the laplacian on $S$, respectively
(subscripts 1 and 2 refer to the first and second 
factor of the $AdS\times S$ manifold).

Using this notation we find from eqs. \t\ the following linearized coupled 
equations for the $(h_2, b)$ scalars
\eqn\set{\eqalign{
& \biggl (\sq_1 +\sq_2 -  {2(p+1)(D-p-3)\over {(D-2)}} e^2 \biggr ) h_2 
- {4(p+2)(D-p-3)\over {(D-2)}} \sq_2  b = 0,\cr
&(\sq_1  + \sq_2) b +  {(p+1)\over{(p+2)}} e^2 h_2= 0.}}
These equations can be easily diagonalized. 
Introducing a complete orthonormal set of
scalar spherical harmonics $Y^I$ satisfying
$\sq_2 Y^I=- \bar e^2 k(k+p+1)Y^I$, where $k$ 
is a non-negative integer depending on the index $I$,
and the harmonic expansions
\eqn\o{h_2(x) = \sum_{I} h_2^I(z) Y^I(y), \ \ \ \ 
b(x) = \sum_{I}  b^I(z) Y^I(y), }
we find that the linear combinations  
\eqn\diag{\eqalign{
s^I &= {1 \over  (2k+p+1)} \biggl ( {1\over 2 (p+2) (D-p-3)} h_2^I 
+ {(k+p+ 1)\over (p+1)(D-2)} b^I \biggr ), \cr
t^I & = {1 \over (2k+p+1)} \biggl ( {1\over 2 (p+2) (D-p-3)} h_2^I 
-{k\over (p+1)(D-2)} b^I \biggr )\cr
}}
obey the equations of motion
\eqn\emef{\eqalign{ \sq_1 s^I &= m_{sI}^2 s^I, \cr
\sq_1 t^I &=m_{tI}^2 t^I, \cr}} 
with masses given by
\eqn\masses{\eqalign{
m_{sI}^2 & = \bar e^2 k(k-p-1), \cr
m_{tI}^2 & = \bar e^2 (k+p+1)(k+2p+2) .\cr}}
We have chosen a suitable 
normalization of the diagonal fields 
such that the inverse relations are simple:
\eqn\inverse{\eqalign{  
h_2^I & =2  (p+2) (D-p-3) \Bigl (ks^I + (k+p+1)t^I\Bigr ),\cr    
b^I &=(p+1)(D-2) ( s^I-t^I ). \cr}}
As a check, we notice that for this particular set of scalar fields 
we have reproduced the masses worked out in refs.
\ref\adsq{
 B. Biran, A. Casher, F. Englert, M. Rooman and P. Spindel,
``The Fluctuating Seven Sphere in Eleven Dimensional Supergravity'',
Phys. Lett. {\bf 134B} (1984) 179\semi
L. Castellani, R. D'Auria, P. Fr\'e, K. Pilch and P. van Nieuwenhuizen, 
``The bosonic Mass Formula for Freund--Rubin Solutions of
$d=11$ Supergravity on General Coset Manifolds'', 
Class. Quantum Grav. {\bf 1} (1984) 33.} 
for the $AdS_4\times S^7$ compactification, and in ref.
\ref\adsset{
P. van Nieuwenhuizen, ``The Complete Mass Spectrum of $d=11$ Supergravity
Compactified on $S^4$ and a General Mass Formula 
for Arbitrary Cosets $M_4$'', Class. Quantum Grav. {\bf 2} (1985) 1.
} for the $AdS_7\times S^4$ one.
The fields $s^I$ describe the Kaluza-Klein tower
of $AdS$ scalars with lowest mass.
To linear order they correspond to the chiral primary operators in SCFT, 
and they will be the focus of our analysis throughout the rest of 
this paper. 
The scalars $t^I$, on the other hand, correspond to descendents of these
chiral primary operators, and for the present purposes 
they can be set to zero.

Having obtained some insights into the diagonal fields $s^I$, we now derive
their quadratic off-shell action starting from eq. \d. 
Proceeding as described in refs. \BZ\ref\AF{
G. Arutyunov and S. Frolov, ``Quadratic Action for type IIB
Supergravity on $AdS_5 \times S^5$'', {\tt hep-th/9811106}.},
one can decouple the $(h_2, b)$ sector 
from $H_{\mu\nu}$ by parameterizing the latter as
\eqn\para{ H_{\mu\nu} = \phi_{\mu\nu} + \nabla_\mu \nabla_\nu \zeta
+ {1\over (D-p-2)} g_{\mu\nu} \eta .}
Requiring a decoupling at the quadratic level between $(h_2, b)$ 
and $ \phi_{\mu\nu}$ fixes $\zeta$ and $\eta $ appropriately. 
At higher orders the decoupling can be dealt with field redefinitions.
Thus, one can immediately 
proceed to identify the quadratic $s^I$ off-shell action
together with their cubic self-couplings.
A laborious calculation produces the following action
(with $d= D-p-3$) 
\eqn\cubicaction{
S = {n\over 2\kappa^2}  \int_{AdS} \!\!\!\! \! 
d^{d+1}z \sqrt{-g_1} 
\biggl [ {1\over 2} \sum_{I}
A_I   s^I (\sq_1 - m_{I}^2) s^I + \sum_{I_{1}I_{2}I_{3}}
{1\over3} G_{I_{1}I_{2}I_{3}} s^{I_1} s^{I_2} s^{I_3}
\biggr ] }
where
\eqn\norm{\eqalign{ n &= (D-2)(p+1)(D-p-3)^2 \cr
A_{I} & = {k (k-1) (2k+p+1)\over (D-p-3) k+ (p+1) } z_I \cr
m_{I}^2 & = \bar e^2 k(k-p-1) \cr
\cr}}
and cubic couplings given by
\eqn\ggeneral{ \eqalign{
&G_{I_{1}I_{2}I_{3}}
 =   {\bar e^2 (D-p-3)\over 2 (p+1)}
\Bigl ( \prod_{i=1}^3 {\alpha_i \over (D-p-3)k_i +p+1} \Bigr)
\Bigl ( 4\alpha^2 - (p+1)^2 \Bigr) \cr
& \biggl [
\Bigl (
(D-2) (2\alpha)^2  -(p+1)(D-2p-4)2\alpha -2 (p+1)^2 
+ 2\Theta (k_1 k_2 +k_2 k_3 +k_3 k_1) 
\Bigr) \cr &
\times \Bigl ( (D-2)2\alpha - (D-2p-4)(p+1) \Bigr )
-4 \Theta [1-p(D-p-4)] k_1 k_2 k_3
\biggr ] a_{I_{1}I_{2}I_{3}} \cabc
\cr}}
with
\eqn\theta{\Theta = (p+1)(D-p-3) -2(D-2).}
The modes with $k=0,1$ decouple from the action
and correspond to some global gauge degree of freedom.
In obtaining these results we have used various definitions for the spherical 
harmonics, similar to those employed in \LMRS. Namely, we describe 
the $n$-sphere of radius $\rho = \bar e^{-1}$ by
$ S^n \equiv \{ \vec y^{\ 2} = \rho^2  \ |\  \vec y  \in R^{n+1} \}$, 
and use scalar spherical harmonics defined by $Y^{I} =
\CC^I_{i_1...i_k}x^{i_1}...x^{i_k}$, 
where the coordinates $x^i = {\bar e y^i}$  live on the unit sphere
and the tensors $\CC^I_{i_1...i_k}$ form
an orthonormal basis for completely symmetric 
traceless tensors. Orthonormality reads as 
$\CC^I_{i_1...i_k} \CC^{J,}{}^{i_1...i_k} = \delta^{IJ}$,
while   $\cabc$ denotes the unique $SO(n+1)$ scalar contraction
of three tensors  $\CC^I_{i_1...i_k}$.
We need to compute
\eqn\sphcont{\eqalign{
\int_{S^n} \!\!\!\! \! d^n y {\sqrt g_2} \
Y^{I} Y^{J} &= z_{I} \delta^{IJ},
\cr
\int_{S^n} \!\!\!\! \! d^n y {\sqrt g_2} \
 Y^{I_1} Y^{I_2} Y^{I_3} &= a_{I_1 I_2 I_3} \cabc,
}}
and obtain 
\eqn\coeff{\eqalign{
z_I & =  \omega_n {(n-1)!! k! \over{(2k +n -1)!!}} \cr
a_{I_1I_2I_3} & = \omega_n {(n-1)!! \over{( 2 \alpha +n -1)!!}}
    {k_1!k_2!k_3! \over \a_1! \a_2! \a_3!}  \cr}}
where $\a_1=\half(k_2+k_3-k_1)$, and so on in a cyclically symmetric fashion,
$\a=\a_1+\a_2+\a_3$ and $\omega_n$  the volume of the sphere
\eqn\vol{
\omega_n= \int_{S^n} \!\!\!\! \! d^n y {\sqrt g_2} =
{2 \pi^{n+1\over 2}\over \Gamma({ n+1\over2})} \Bigl ( {1\over
\bar e} \Bigr )^n .}

Returning to the values of the cubic couplings 
$G_{I_{1}I_{2}I_{3}}$ given in eq. \ggeneral,
we note that they simplify for the 
$AdS_{4,7,5}\times S^{7,4,5}$ cases, since then the coefficient
$\Theta=0$.
In particular, defining $G_{I_{1}I_{2}I_{3}} = 
\bar G_{I_1 I_2 I_3} a_{I_1 I_2 I_3} \cabc $,
 we obtain
\eqn\gqua{\eqalign{ AdS_4 \times S^7: \ \ \ \ \ \ \ \
& \bar G_{I_1 I_2 I_3}  =  24 \bar e^2 \Bigl (\prod_{i=1}^3
{\alpha_i \over k_i+2}\Bigr) (\alpha +2) ( \alpha^2 -1)(\alpha^2 -9) \cr
AdS_7 \times S^4: \ \ \ \ \ \ \ \
& \bar G_{I_1 I_2 I_3} =  3 \bar e^2 \Bigl (\prod_{i=1}^3
{\alpha_i \over 2k_i+1}\Bigr)
 (2\alpha -2) (4\alpha^2 -1)(4\alpha^2 -9) \cr
AdS_5\times S^5: \ \ \ \ \ \ \ \
& \bar G_{I_1 I_2 I_3} 
=   16\bar e^2 \Bigl (\prod_{i=1}^3{\alpha_i \over k_i+1}  \Bigr)
\alpha (\alpha^2 -1)(\alpha^2 -4).}}

We are now ready to compute two and three point functions
in the SCFTs using the $AdS$/CFT correspondence.
The general formulas derived in 
\ref\FMR{D. Freedman, S. Mathur, A. Matusis and L. Rastelli,
``Correlation functions in the CFT(d)/AdS(d+1) correspondence'',
Nucl. Phys. {\bf B546} (1999) 96, {\tt hep-th/9804058}.} and 
adapted to the action \cubicaction\
give (with $d= D-p-3$ and $AdS$ radius set to 1)
\eqn\twoptf{
\langle \CO^{I}(x) \CO^{J}(y) \rangle = 
{ n A_I \over 2 \kappa^2}  {2\Delta - d \over 
\pi^{d\over2}} {\Gamma(\Delta)\over \Gamma(\Delta -{d\over2})}
{(w^I)^2 \delta^{IJ} \over |x-y|^{2\Delta}}
}
and 
\eqn\threeptf{
\langle \CO^{I_1}(x) \CO^{I_2}(y) \CO^{I_3}(z) \rangle = 
{R_{I_1I_2I_3}
\over |x-y|^{\Delta_1+\Delta_2-\Delta_3}
|y-z|^{\Delta_2+\Delta_3-\Delta_1}
|z-x|^{\Delta_3+\Delta_1-\Delta_2}},}
where
\eqn\threeptff{\eqalign{
R_{I_1I_2I_3} 
= &  \Bigl ({ n\over 2 \kappa^2}\Bigr ) 
{1\over \pi^d}
{\Gamma({1\over2}(\Delta_1+\Delta_2-\Delta_3))
\Gamma({1\over2}(\Delta_2+\Delta_3-\Delta_1))
\Gamma({1\over2}(\Delta_3+\Delta_1-\Delta_2))
\over 
\Gamma(\Delta_1 -{d\over2})
\Gamma(\Delta_2 -{d\over2})
\Gamma(\Delta_3 -{d\over2})} \cr &
\Gamma(\hbox{${1\over2}$}({\Delta_1+\Delta_2+\Delta_3} - d))
G_{I_1I_2I_3} 
{w^{I_1}w^{I_2}w^{I_3} }.
}}
The factors $w^{I}$ parameterize unknown proportionality constants
which relate the fields $s^I$ to the sources of the operators
$\CO^I$, as in \LMRS.
These factors can presumably be fixed by carefully 
studying absorption processes on the branes \ref\kleb{
I. Klebanov, ``World Volume Approach to Absorption by Non-dilatonic Branes'',
Nucl. Phys. {\bf B496} (1997) 231, {\tt hep-th/9702076}\semi
S. Gubser and I. Klebanov,
``Absorption by Branes and Schwinger Terms in the World Volume Theory''
Phys. Lett. {\bf B413} (1997) 41, {\tt hep-th/9708005}.}.
However, for the present purposes we follow ref. \LMRS\
and fix them to normalize the two point functions as
\eqn\twoptff{
\langle \CO^{I}(x) \CO^{J}(y) \rangle = 
{\delta^{I J} \over |x-y|^{2\Delta}}.}

We first study the case of $AdS_4 \times S^7$, which corresponds
to SCFT$_3$ at large $N$ and constitutes our main interest.
To match the  $AdS_4 \times S^7$ solution to the near horizon geometry
of $N$ M2 branes we tune the  gravitational coupling
${1\over 2\kappa^2} = {N^{3\over 2} \over 2^{8} {\sqrt 2} \pi^5}$.
Also, to employ the formulas given above, we need to fix the $AdS$ radius
to one by setting the mass parameter $\bar e={1\over 2}$. 
The masses of the scalars $s^I$ fix the conformal dimensions of the 
operators $\CO^I$ to be $ \Delta_I= {k\over 2}$. We obtain
\eqn\rtre{  R_{I_1I_2I_3} = { \pi \over N^{3\over 4}} 
{2^{-\alpha -{1\over 4}} \over \Gamma({\alpha\over 2}+1)}
\biggl ( \prod_{i=1}^3 { \sqrt{\Gamma(k_i +2)}
\over \Gamma({\alpha_i+1 \over 2})}\biggr ) \cabc.}
This is the $AdS$/CFT prediction for the normalized 
CPO three point functions in SCFT$_3$ at large $N$.

Our general formulation can also produce the CPO three point functions 
in SCFT$_6$. In fact, the Chern-Simon terms doesn't give any contribution
in this case as well, and it can be neglected from the start.
Thus, we turn immediately to the $AdS_7 \times S^4$ case and
set the $AdS_7$ radius to one by fixing
${1\over 2\kappa^2} = {4 N^3 \over \pi^5}$ and $\bar e=2$. 
The  masses of the scalars $s^I$ fix the conformal dimensions of the 
operators $\CO^I$
to be $ \Delta_I= 2k$ so that
\eqn\rdue{  R_{I_1I_2I_3} = {2^{2\alpha -2}\over (\pi N)^{3\over 2}} 
\Gamma(\alpha) \biggl (
\prod_{i=1}^3 {\Gamma(\alpha_i + {1\over 2})\over 
\sqrt{\Gamma(2 k_i -1)}} \biggr ) \cabc.}
This essentially confirms the results already obtained in \CFM\
(apart from a factor $\pi^{1\over4} 2^{{\alpha -11\over 2}}$ 
present there for which we cannot find agreement). 
The three point couplings for the $s^I$ scalars at level $k=2$
(and the corresponding three point functions)
could in principle be extracted also from ref.
\ref\PVN{ H. Nastase, D. Vaman and P. van Nieuwenhuizen,
``Consistent nonlinear KK reduction of 11d supergravity on $AdS_7\times S^4$ 
and self-duality in odd  dimensions'',
{\tt hep-th/9905075}.},
where the complete consistent Kaluza-Klein reduction of
11$D$ supergravity on $AdS_7\times S^4$ has been worked out.
However, it is not straightforward to extract the cubic couplings
from the lagrangian given there, since one should first perform 
some gauge-fixing to identify the physical scalars $s^I$.

Finally, we consider our model on $AdS_5\times S^5$. 
Using the $AdS$/CFT correspondence we can obtain correlation
functions for the operators $\CO^I$ corresponding to 
the scalars $s^I$. The result will not be obviously related to 
$d=4$, $\CN=4$ SYM theory, since in the latter case the sources for the 
dual gravitational model must be self-dual $3$-branes, 
i.e. branes with both electric and magnetic charges.
Nevertheless, to satisfy our curiosity we compute the correlation functions.
We set the $AdS$ radius to 1 by fixing $\bar e=1$.
We also set the gravitational coupling to
${1\over 2\kappa^2} = {N^2 \over 4 \pi^5}$, as would have been appropriate 
to the near horizon geometry of $N$ D3 branes.
This time the $s^I$ masses fix the conformal dimensions of the operators
to be $ \Delta_I= k$ and we get 
\eqn\runo{  R_{I_1I_2I_3} = {1\over N} \sqrt{k_1k_2k_3}\ \cabc.}
Surprisingly, this reproduces the $d=4$, $\CN=4$ SYM  
results at large $N$ obtained in \LMRS. 
We have no immediate explanation for this coincidence.

To conclude, we have used the $AdS$/CFT conjecture to predict the 
normalized three point correlation functions of the chiral 
primary operators in SCFT$_3$ and SCFT$_6$. 
The first result is new, while the latter essentially
confirms the findings of ref. \CFM.
Adapting our calculation to the $AdS_5\times S^5$ case, we have noticed 
that our model reproduces the normalized three point functions
at large $N$ for the $d=4$, $\CN=4$ SYM theory.

We plan to present more details of our computation in a future
publication, possibly producing additional correlation functions
involving scalar operators related to the $t^I$ fields.

\listrefs
\end